# A Search Procedure for Cyclic Subsets

Paul Clarke

St. Paul's College Raheny

January 5[th] 2014

**Abstract**

In this paper, a polynomial time algorithm for finding the set of all cyclic subsets in a graph is presented. The concept of cyclic subsets has already been introduced in an earlier paper. The algorithm finds cyclic subsets in a graph $G$ by conjoining "building block" subsets of length three in $V(G)$. We prove the correctness of this algorithm and present an asymptotic time complexity analysis of the algorithm's performance.

## 1. Introduction

Cyclic subsets were first introduced in formal graph theory in the following paper on Barnette's conjecture: [1]. A cyclic subset in a graph $G$ is defined as a subset of $V(G)$ such that the subgraph induced by that subset is a cycle. We see how cyclic subsets are fundamental to graphs, and in particular, Hamiltonian graphs.

In section two of this paper, basic graph theoretical terminology is defined. In section three, we present the algorithm in pseudocode along with an informal breakdown of the algorithm. In section four, we prove rigorously the correctness of the algorithm. We conclude this paper with an analysis of time complexity, confirming that the time complexity function of the algorithm is polynomial.

## 2. Terminology

In this section, graph theoretical terminology is defined for use in this paper. For graph theoretical terminology not defined here, consult the books of Harary [2] and Bondy [3]. Computational complexity theory terminology will not be defined here, so consult the books of Sipser [4] and Garey [5] for the relevant material.

We define a graph $G = (V, E)$, where $V$ is a finite non-empty set called the *vertex set*. The elements of $V(G)$ are called *vertices*. The elements of $E(G)$ are two element subsets of $V(G)$ and are called *edges*. In general, $n = |V(G)|$. Two vertices $v, u \in V(G)$ are *adjacent* in $G$ if and only if $\{v, u\} \in E(G)$. The number of vertices adjacent to a vertex $v$, denoted $deg(v)$, is the *degree* of $v$.

The *complement* $G'$ of a graph $G$ is a graph whereby $V(G') = V(G)$ and $\{v_i, v_j\} \in E(G')$ if and only if $\{v_i, v_j\} \notin E(G)$. A graph $G' = (V, E)$ is a subgraph of $G$ (denoted $G' \subseteq G$) if $V(G') \subseteq V(G)$ and $E(G') \subseteq E(G)$. A *walk* of length $k$ in a graph $G$ from vertex $v$ to vertex $u$ is a subgraph $W \subseteq G$ whereby $V(W)$ is a sequence of vertices $\{v = v_0, v_1, v_2, ... , v_k = u\}$ in $G$ such that $\{v_i, v_{i+1}\} \in E(G)$ for all $0 \leq i < k$. $E(W)$ is defined as $\{\{v_i, v_{i+1}\} \mid 0 \leq i < k, v_i \in V(W)\}$. The two vertices $v_1, v_2 \in V(W)$ whereby $deg(v_1) = 1$ and $deg(v_2) = 1$ are called the *endpoints* of $W$.

A *path* $P$ is defined as a walk whereby for every vertex $v \in V(P)$, $deg(v) \leq 2$. A *cycle* $C$ in a graph is a path with an extra edge connecting the two endpoints of the path. A *k-cycle* is a cycle $C$ whereby $|V(C)| = k$. If there exists a cycle $C$ in $G$ such that $|V(C)| = |V(G)|$, $G$ is *Hamiltonian*. The spanning cycle is called a *Hamiltonian cycle* in $G$.

*A vertex induced subgraph* in a graph $G$ is a subgraph $G' \subseteq G$ with a vertex set $V(G') \subseteq V(G)$ and an edge set $E(G')$ consisting of every edge $\{v_i, v_j\} \in E(G)$ such that $v_i \in V(G')$ and $v_j \in V(G')$. A *cyclic subset $C$* in $G$ is a subset of vertices in $G$ such that $S$ is a cycle, where $S$ is the subgraph induced by $C$ in $G$. We define $S(G)$ as the set of all cyclic subsets in a graph $G$.

In this paper, we consider only finite, simple, undirected graphs.

## 3. Algorithm breakdown

In this section, I present my algorithm for finding the set of all cyclic subsets $S(G)$ in a graph $G$. The algorithm is based around the concept of "building blocks" of length three in graphs. The algorithm is called cycsub. The algorithm cycsub can be seen in pseudocode in Listing 1.

Listing 1: The "cycsub" algorithm

cycsub($V(G)$, $E(G)$):

1. Take the set $X$ of all subsets $S$ of $V(G)$ such that $|S| = 3$

2. Let $F(G) = \{\}$
   Let $T(G) = \{\}$
   Let $Q(G) = \{\}$

   for $i$ in $X$:
       Let $S'$ be the subgraph induced by $i$ in $G$
       Let $S^*$ be the complement graph of $S'$
       if $|E(S')| = 1$:
           Append $(i, E(S^*))$ to $T(G)$
       else if $|E(S')| = 2$:
           Append $(i, E(S^*))$ to $F(G)$
       else if $|E(S')| = 3$:
           Append $i$ to $Q(G)$

3. Let $I(G) = F(G)$
   Let $Z = \{\}$

   while $I(G)$ is nonempty:
       for $i$ in $I(G)$:
           Let count $= 0$
           for $j$ in $Q(G)$:
               if $j \subseteq i$:
                   count $=$ count $+ 1$
           if count $> 0$:
               Remove $i$ from $I(G)$

  for $i$ in $I(G)$:

    for $j$ in $F(G)$:

      if $i[1] = j[1]$ and $j[0] \not\subset i[0]$:

        Append $i[0] \cup j[0]$ to $Z$

  Let $I'(G) = I(G)$

  Let $I(G) = \{\}$

  for $i$ in $I'(G)$:

    for $j$ in $T(G)$:

      if $|i[1] \cap j[1]| = 1$ and $j[0] \not\subset i[0]$:

        Append $(i[0] \cup j[0], j[1] / (i[1] \cap j[1]))$ to $I(G)$

4. for $i$ in $Q(G)$:

  Append $i$ to $Z$

5. Let $Z' = \{\}$

  for $i$ in $Z$:

    Let count $= 0$

    for $j$ in $Z$:

      if $j \subseteq i$ and $i \neq j$:

        count $=$ count $+ 1$

    if count $= 0$:

      Append $i$ to $Z'$

6. return $Z'$

  I now provide an informal breakdown of cycsub. The purpose of this section is to give less technical readers a general insight into how the algorithm works.

Step 1:

  In this step, all subsets of length 3 in $V(G)$ are taken and appended to a set $X$. Clearly, $|X| = {_n}C_3$, where $n = |V(G)|$.

Step 2:

In this step, the subsets of $V(G)$ taken in step 1 are individually analyzed and classified into three different sets based on the number of edges in the subgraphs induced by these subsets. In order, these sets are $F(G)$: *foundations of G*, $T(G)$: *extensions of G*, and $Q(G)$: *3-cliques of G*.

Step 3:

This is the most important step in the algorithm. In this step, cyclic subsets are formed by repeatedly joining extensions with foundations. We use the set $I(G)$ as a processing ground for forming cyclic subsets.

In each iteration of the while loop, elements of $I(G)$ join with extensions to form paths of vertices and edges. These paths are eventually closed off with foundations to form cycles. The power of this step is in the restriction on the number of such cycles that can be formed and appended to $Z$.

Step 4:

By our definition of a cyclic subset, all elements of $Q(G)$ must be cyclic subsets. Therefore, in this step, every element of $Q(G)$ is appended to $Z$.

Step 5:

In this step, for every element $C$ in $Z$, we check if any other elements in $Z$ are subsets of $C$. If so, $C$ clearly cannot be a cyclic subset, and so $C$ is removed from $Z$.

Step 6:

As will be shown later, $Z$ is now the set of all cyclic subsets in $G$, and this is the set returned by cycsub.

## 4. Proof of correctness

In this section, we show that cycsub($V(G)$, $E(G)$) returns the complete set of cyclic subsets in a graph $G$. In other words, we wish to prove that cycsub($V(G)$, $E(G)$) = $S(G)$.

**Theorem 1:** If $G$ is a graph, then $S(G)$ = cycsub($V(G)$, $E(G)$).

**Proof:** Consider the algorithm cycsub in Listing 1. $X$ consists of all subsets of length 3 in $V(G)$. As described in section 3, these subsets are classified into 3 sets as in Listing 1: $F(G)$, $T(G)$ and $Q(G)$. For $i$ in $F(G)$ or $T(G)$, $i[1]$ is the set of all nonadjacent pairs of vertices in $S$, where $S$ is the subgraph induced by $i$ in $G$.

Consider step 3. We let $I(G) = F(G)$ before the first loop of the while loop $W$ in step 3. We also initiate the set $Z$. As can be seen in $W$, a new $I(G)$ is formed after each loop of $W$. We define $I_m(G)$ as $I(G)$ after the $m^{th}$ loop of $W$. Similar to the above, for $i$ in $I_m(G)$, $i[1]$ is the set of all nonadjacent pairs of vertices in $S$, where $S$ is the subgraph induced by $i$ in $G$.

Now consider some cyclic subset $C \in S(G)$ whereby $|C| > 3$. Consider some set of three consecutive vertices $\{v_i, v_j, v_k\}$ in $C$. (Fig. 1) As $C$ is a cyclic subset, $\{v_i, v_j, v_k\} = R[0]$ for some $R \in F(G)$. Therefore, $R \in I_0(G)$. Suppose that $|C| = 4$. We define $F_2$ as the second for loop in $W$. Let $v_x \in C$ such that $v_x \notin R[0]$. Clearly, $v_x$ is unique as $|C| = 4$.

As $C$ is a cyclic subset, $\{v_x, v_i, v_k\} = L_1[0]$ for some $L_1 \in F(G)$. In $F_2$, for $i$ in $I_0(G)$, and for $j$ in $F(G)$, we check if $i[1] = j[1]$ and $i[0] \cap j[0] \neq j[0]$. As $L_1 \in F(G)$ and $R \in I_0(G)$, we can let $i = R$ and $j = L_1$. Clearly, $R[1] = L_1[1]$ and $L_1[0] \not\subset R[0]$, as the subgraphs induced by $R[0]$ and $L_1[0]$ clearly share a common pair of nonadjacent vertices: $\{v_k, v_i\}$ Then, $C = R[0] \cup L_1[0] = \{v_x, v_i, v_k, v_j\}$ is appended to $Z$.

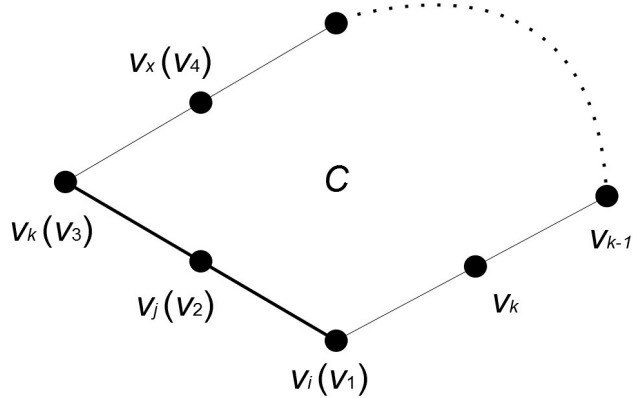

Fig. 1: The subgraph induced by A formed by applying cycsub to a cyclic subset C

Now suppose that $|C| > 4$. Once again consider $R$. We let $F_3$ be the third for loop in $W$. In $F_3$, for $i$ in $I_0'(G) = I_0(G)$, and for $j$ in $T(G)$, we check if $|i[1] \cap j[1]| = 1$ and $i[0] \cap j[0] \neq j[0]$. Consider, $\{v_i, v_k, v_x\}$, where $v_x$ is the next consecutive vertex in $C$. As $C$ is a cyclic subset, $\{v_i, v_k, v_x\} = L_2[0]$ for some $L_2 \in T(G)$. As $R \in I_0(G)$ and $L_2 \in T(G)$, we can let $i = R$ and $j = L_2$. Clearly, $|R[1] \cap L_2[1]| = 1$ and $L_2[0] \not\subset R[0]$, as the subgraphs induced by $R$ and $L_2$ share a common pair of nonadjacent vertices: $\{v_k, v_i\}$. Then, $A = (\{v_i, v_j, v_k, v_x\}, \{v_i, v_x\})$ is appended to $I_1(G)$, and $W$ repeats with $I_1(G)$ in the place of $I_0(G)$.

We let $k = |C|$. The above process will now repeat with $A$ in the place of $R$, as there will always be some $L_2 \in T(G)$ in $C$ such that $L_2$ will join with $A$ in $F_3$, as described above, up until $|A[0]| = k - 1$. (Fig. 1) Note that in subsequent $I_m(G)$, only vertices on the endpoints of $S$ are considered as nonadjacent vertices, where $S$ is the subgraph induced by any arbitrary element $i \in I_m(G)$. In $I_{k-4}(G)$, we let $A[0] = \{v_1, v_2, v_3, v_4 \ldots v_{k-1}\}$ in consecutive order with respect to $C$. By the above process, $A[1] = \{v_1, v_{k-1}\}$. By this definition, $C / A[0] = \{v_k\}$.

As $C$ is a cyclic subset, for some $L_3 \in F(G)$, $\{v_1, v_k, v_{k-1}\} = L_3[0]$. Then, in $F_2$, as $L_3[1] = \{v_1, v_{k-1}\} = A[1]$, and $L_3[0] \not\subseteq A[0]$, $C = L_3[0] \cup A[0]$ is appended to $Z$, as was described above. Now suppose that $|C| = 3$. Then $C \in Q(G)$. By step 4 of cycsub, $C$ is then appended to $Z$.

By the above constructions, we can deduce that if $C$ is a cyclic subset, then $C \in Z$ after step 4 of cycsub. Consider step 5 of cycsub. In this step, $Z'$ is initiated. For $i$ in $Z$, and for $j$ in $Z$, we check if $j \subseteq i$ and $i \neq j$. If this is true for any $j$, $i$ is not appended to $Z'$. Clearly, by the constructions above, for any $i \in Z$, $S$ is a Hamiltonian graph, where $S$ is the subgraph induced by $S$ in $G$.

Consider the case where $i = C$, where $C$ is some arbitrary cyclic subset in $Z$. Let $S$ be the subgraph induced by $j$ in $G$. Suppose that $j = C$. Then the above condition is not satisfied, as $i \neq j$. Now suppose that $j \neq C$. Then $j \not\subseteq i$ as no Hamiltonian graph can be a subgraph of the graph induced by a cyclic subset that is not itself the induced subgraph itself.

Therefore, if $C$ is a cyclic subset, $C \in Z'$. Suppose now that for some $i \in Z'$, $i$ is not a cyclic subset. Let $H$ be the Hamiltonian cycle in $S$, where $S$ is the subgraph induced by $i$. As $i$ is not a cyclic subset, there must exist some edge $e \in E(S)$ such that $e$ is not on $H$. However, if $G$ is a Hamiltonian graph, and if there exists some edge $e$ such that $e$ is not on $H$, there must exist some cyclic subset $c \in S(G)$ such that $c \subseteq i$. Therefore, $C$ will be removed from $Z'$ in step 5, implying that $i \notin Z'$, a clear contradiction. Therefore, if $i \in Z'$, $i$ is a cyclic subset.

Therefore, $C \in Z'$ if and only if $C$ is a cyclic subset in $S(G)$. Therefore, if $G$ is a graph, $S(G) = Z' = $ cycsub$(V(G), E(G))$ by step 6 of cycsub, and the theorem follows. QED

## 5. Time complexity

In this section, we show that cycsub is a polynomial time algorithm. The input length $n$ for cycsub will be measured in terms of the number of vertices in $G$, or $|V(G)|$. $|V(G)|$ and $|E(G)|$ can only differ by a constant factor, so this has no effect on the time complexity function of cycsub.

For the purposes of clarity, we assume that the operation of appending an element to a set requires $n$ elementary operations. Similar to the above, this definition has no effect on the overall time complexity of the algorithm.

**Theorem 2:** The time complexity function of cycsub is $O(n^9)$.

**Proof:** We analyse each step of cycsub individually.

Step 1:

For all $_nC_3$ subsets $S$ of length 3 in $V(G)$, $S$ is appended to $X$, where $n = |V(G)|$. If each subset takes $n$ elementary operations to append to $X$, then the time complexity of this step is clearly $_nC_3(n) = O(n^4)$.

Step 2:

In this step, for every element in $X$, an if statement is computed. Clearly, this if statement can be computed in linear time. For each of these if statements, an element is appended to a set. Therefore, as $|X| = {_nC_3}$, the time complexity of this step is $_nC_3(n + n) = O(n^4)$.

Step 3:

We require the following lemma to accurately establish the time complexity of this step.

**Lemma 1:** For any $m$, $|I_m(G)| \leq |F(G)|$.

**Proof:** Suppose that for some $m$, we have $I_m(G)$. We are required to prove that for every element $e \in I_m(G)$, there exists a unique element $f \in F(G)$. Consider some $e = (\{v_1, v_2, v_3, \ldots, v_{m+3}\}, \{v_1, v_{m+3}\}) \in I_m(G)$. Consider $f = (\{v_{m+1}, v_{m+2}, v_{m+3}\}, \{v_{m+1}, v_{m+3}\}) \in F(G)$. (Fig. 2)

Suppose that $f$ is not unique to $e$. Therefore, there exists some element $e' = (\{v_a, v_b, \ldots v_{m+1}, v_{m+2}, v_{m+3}\}, \{v_a, v_{m+3}\}) \in I_m(G)$ such that $f[0] \subseteq e'[0]$. However, as $|e'[0]| > 3$, $f' = (\{v_a, v_b, v_c\}, \{v_a, v_c\}) \in F(G)$, providing a unique foundation for $e'$. Note that $f' \notin Q(G)$ as then $e'$ would be removed from $I_m(G)$ in the first for loop $F_1$ of the while loop $W$ in step 3 of cycsub. (Fig. 2)

We can repeat the above process recursively until for every element $e \in I_m(G)$, we have shown that there exists a unique element $f \in F(G)$. Therefore, for any $m$, $|I_m(G)| \leq |F(G)|$. QED

In the worst case, $W$ can loop $n - 3$ times, where $n = |V(G)|$. This is because in $I_m(G)$, for every element $e \in I_m(G)$, $|e[0]| = m + 3$. In the worst case, $|F(G)| \approx {_nC_3}$. Therefore, by Lemma 1, in the worst case, $|I_m(G)| \approx {_nC_3}$. For every loop of $W$, and for every element in $I_m(G)$, $F_1$ iterates through

$F(G)$ and considers an if statement. Then, an element is appended to $Z$.

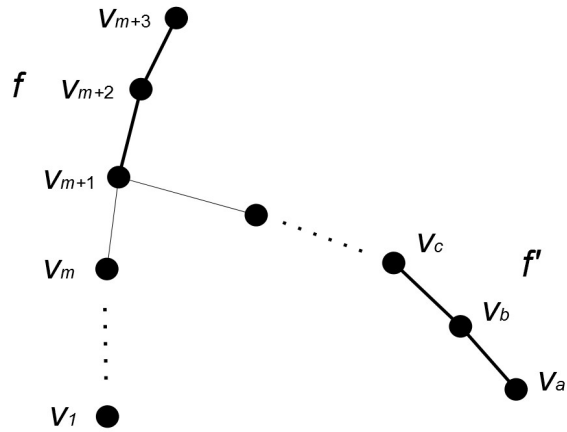

Fig. 2: For every element $e \in I_m(G)$, there exists a unique element $f \in F(G)$

Assuming that this if statement requires $n$ elementary operations to compute, and appending an element to a set requires $n$ elementary operations, the time complexity of $F_2$ in the worst case is $(n - 3)(_nC_3)(_nC_3)(n + n) = O(n^8)$. Clearly, $F_3$ has the same time complexity function. Therefore, the overall time complexity of this step is $2(O(n^8)) = O(n^8)$.

Step 4:

In the worst case, $|Q(G)| \approx {_nC_3}$. Assuming that appending an element to a set requires $n$ elementary operations, the time complexity of this step is $(_nC_3)(n) = O(n^4)$.

Step 5:

In the worst case, for every iteration of $W$ in step 3, $_nC_3$ elements can be appended to $S(G)$. Therefore, in the worst case, $|Z| = (n - 3)(_nC_3)$ in step 5. Incrementing the variable count is negligible. Therefore, as there are two nested for loops, the time complexity of this step is $(n - 3)(_nC_3)(n - 3)(_nC_3)(n) = O(n^9)$.

Step 6:

This step runs in linear time, or $O(n)$ time complexity.

Therefore, the overall time complexity function of cycsub is $O(n^4) + O(n^4) + O(n^8) + O(n^4) +$

$O(n^9) + O(n) = O(n^9)$. QED